\documentclass[a4paper]{article}
\usepackage{amsfonts}
\usepackage{latexsym}
\usepackage{epsfig}
\def\b{\bar}
\def\d{\partial}

\def\cA{{\cal A}}
\def\cD{{\cal D}}

\def\m{\mu}
\def\n{\nu}

\def\t{\tau}

\def\~{\tilde}
\def\h{\eta}

\def\bY3{\bar Y_{,3}}
\def\Y3{Y_{,3}}
\def\z{\zeta}
\def\Z{{\b\zeta}}
\def\Y{{\bar Y}}
\def\cZ{{\bar Z}}
\def\`{\dot}
\def\be{\begin{equation}}
\def\ee{\end{equation}}
\def\bea{\begin{eqnarray}}
\def\eea{\end{eqnarray}}

\def\fn{\footnote}

\def\cF{{\cal F}}

\def\mn{{\mu\nu}}

\begin{document}

\title{Orientifold and stringy frame of the Kerr Spinning Particle}

\author{Alexander Burinskii\\
NSI Russian Academy of Sciences\\
B. Tulskaya 52, 115191 Moscow, Russia}
\maketitle

\begin{abstract}
\noindent
We show that the frame of the Kerr spinning particle consists of two
topologically coupled strings.  One of them is
the Kerr singular ring representing a string with an
orientifold world-sheet. It has electromagnetic excitations
(traveling waves), which corresponds to the old model
of the Kerr's microgeon. The excitations induce the appearance of the
extra axial string which is topologically coupled to the Kerr circular
string and is a carrier of  pp-waves and fermionic zero modes. This string
can be described by the Witten field model for superconducting strings.
\end{abstract}

\section{ Introduction}
The Kerr rotating black hole solution displays some remarkable
relations to spinning particles
\cite{Car,Bur0,IvBur,BurSen,BurStr,BurSup,Bag,Ori}. For the
parameters of elementary particles, $|a|>> m$, the black-hole
horizons disappear, obtaining the source in the form of a closed
singular ring of the Compton radius $a=J/m$. In the old model of the Kerr
spinning particle \cite{Bur0} this ring was considered as a
gravitational waveguide  containing traveling electromagnetic (and
fermionic) wave excitations. The assumption that the Kerr singular
ring represents a closed relativistic string was advanced about
thirty years ago \cite{IvBur}, which has got confirmation on the level
of the evidences \cite{BurSen,IvBur}. However, the attempts
to show it explicitly ran into obstacles.
The motion of the Kerr ring is the lightlike sliding
along itself, which could be described as a string containing
lightlike modes of only one direction.  However, the
string equations do not admit such solutions.
 This problem can be resolved
by introducing an orientifold world-sheet \cite{Ori}.

Another type of excitations is a vibration of the Kerr circular
string. It is connected with a complex structure of the Kerr geometry
\cite{Bur-nst}. Evolution of the complex Kerr,s retarded-time parameter
$\t =t+\sigma$ forms a world-sheet. It corresponds to
a hyperbolic string which is very similar to the $N=2$ string.
It was shown that consistence of this string demands also the orientifold
world-sheet structure \cite{BurStr}.

The novel feature appears when we consider the exact solutions for
electromagnetic excitations of the Kerr circular string. One obtains
 the appearance of the extra axial string which is connected with the
NUT parameter and is infinite (for a free particle). This string is
related to the Dirac monopole string and is topologically coupled
to the Kerr circular string. It acquires the traveling pp-waves
which are induced by excitations of the Kerr string and can
be described by the Witten field model for superconducting strings.
We conjecture that this string may play the role of a carrier of
the wave function.

Our treatment is based on the Kerr-Schild formalism  \cite{DKS}
and the results of previous paper \cite{Bur-nst} where the real and complex
structures of the Kerr geometry were considered.
\section{ Orientifold structure of the Kerr string}
The Kerr-Schild approach \cite{DKS} is based on the Kerr-Schild
form of the metric: $ g_{\m\n}= \h_{\m\n} + 2 h k_{\m} k_{\n} $
where $ \h_\mn $ is the metric of auxiliary Minkowski space-time,
$ h= \frac{mr-e^2/2}{r^2 + a^2 \cos^2 \theta},$
and $k_\m$ is a twisting
null field, which is tangent to the Kerr principal null congruence
(PNC) which is determined by a twisted 1-form
\fn{The rays of the Kerr PNC
are twistors and the Kerr PNC is determined by the Kerr theorem, see detailed
description in \cite{Bur-nst}.}
The field $k^\m$ is null with respect to $\h_\mn$, as well as with respect to
the full metric $g_\mn$.

  The Kerr singular ring is the branch line of the
Kerr space on two folds: positive sheet ($r>0$) and `negative' one
($r<0$). Since for $|a|>>m$ the horizons disappear, there appears
the problem of the source of the Kerr solution with the
alternative: either to remove this twofoldedness or to give it a
physical interpretation.
\par
The negative sheet can be retained if we shall treat it as the sheet
of advanced fields. In this case the source of the spinning
particle turns out to be the Kerr singular ring with the
electromagnetic excitations in the form of traveling waves, which
generate spin and mass of the particle.  A model of this sort
was suggested long ago as a ``microgeon with spin" \cite{Bur0}, where
the Kerr singular ring was considered as a waveguide providing a
circular propagation of the electromagnetic or
fermionic wave excitations.
The lightlike structure of the Kerr ring is seen from
the analysis of the PNC which contains tangent to the ring lightlike rays.
It is seen from the form of $k^\m$.  Approaching the ring
($r\to 0$) PNC takes the form $ k \simeq
dt - (xdy-ydx)/a = dt -a d\phi $, and the light-like vector field
$k_\m$ turns out to be tangent to the world sheet of the Kerr ring.
It shows
that the Kerr ring is sliding along itself with the speed of the
light. It was recognized long ago \cite{IvBur} that the Kerr singular ring
 can be considered as the string having the traveling wave excitations
in the model of  microgeon.
The Kerr twofoldedness shows that it is the ``Alice" type of string.
One of the evidences of its stringy structure was obtained by the analysis
of the axidilatonic generalization of the Kerr solution
(given by \cite{Sen}).  It was shown \cite{BurSen} that the fields
 near the Kerr ring are similar to the field around a heterotic string.

The world-sheet of the Kerr ring satisfies the
string wave equation and constraints; however, there appear
the problem with boundary conditions.
Representing solution as the sum of the `left' and `right' modes
\begin{equation}
X_R^\m (\t -\sigma) =\frac 12 [ x^\m + l^2 p^\m (\t -\sigma) +
il\sum_{n\ne0} \frac 1n \alpha_n^\m e^{-2in(\t-\sigma)}] ,
\label{R}
\end{equation}
\begin{equation}
X_L^\m (\t +\sigma) =\frac 1 2
[ x^\m + l^2 p^\m (\t +\sigma) +
il\sum _{n\ne0} \frac 1n \tilde\alpha_n^\m e^{-2in(\t+\sigma)}] ,
\label{L}
\end{equation}
one sees that the string constraints $\dot X_\m \dot X^\m + X^{\prime}_\m
X^{\prime\m} =0, \qquad \dot X_\m  X^{\prime\m} =0$
[$()^\prime \equiv \d _\sigma ()$],
are satisfied if the modes are lightlike
$(\d _ \sigma X_{L(R)\m}) (\d _{\sigma} X_{L(R)}^\m) =0.$

Indeed, setting $2\sigma = a \phi$ one can describe the lightlike
worldsheet of the Kerr ring (in the rest frame of the Kerr
particle) by the lightlike surface
\be X_L^\m(t,\sigma) = x^\m + \frac 1{\pi T} \delta _0^\m p^0 (t + \sigma) +
\frac a 2 [(m^\m +in^\m)
e^{-i2(\t+\sigma)} + (m^\m -in^\m) e^{i2(\t+\sigma)}]  ,
\label{kring} \ee
where $m^\m$ and $n^\m$ are two spacelike basis
vectors lying in the plane of the Kerr ring.  One can see that
$\d _\sigma X_L^{\m}$
will be a light-like vector if one sets $p^0 = 2\pi a T $.
\par
Therefore, the Kerr world-sheet could be described by modes of one (say
``left") null direction.
However, one sees that the closed string boundary condition
$ X^\m(\t,\sigma)=X^\m(\t,\sigma+\pi) $
 will not be satisfied since the time component
$X_L^0 (t,\sigma +\pi)$ acquires contribution from the second term
in (\ref{kring}), which is usually compensated by this term from
the `right' mode. Therefore, the strings having only the ``left"
modes cannot be closed.

\begin{figure}[ht]
\centerline{\epsfig{figure=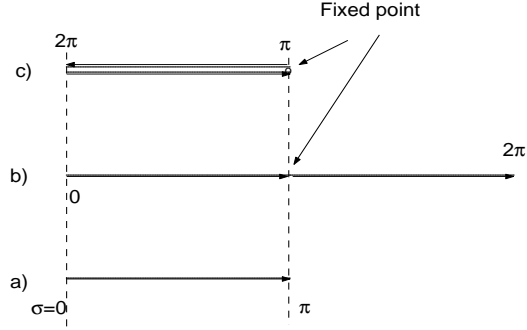,height=5cm,width=7cm}}
\caption{Formation of orientifold: a) the initial string interval;
b) extension of the interval and formation of the both side
movers; c) formation of the orientifold.} \end{figure}

This obstacle can be removed by the formation of the
world-sheet orientifold.
The interval of an open string  $\sigma
\in [0,\pi] $ is formally extended to $[0,2\pi]$, setting
\be X_R (\sigma
+\pi) =X_L (\sigma), \qquad X_L (\sigma +\pi) =X_R (\sigma).  \label{ext} \ee
By such an extension, the both types of modes, ``right" and
``left", appear since the ``left" modes  play
the role of ``right" ones on the extended piece of interval. If
the extension is completed by the changing of orientation on the
extended piece,  $\sigma ^\prime = \pi - \sigma $, with a
subsequent identification of $\sigma$ and  $\sigma ^\prime$, then
one obtains the closed string on the interval $[0,2\pi]$ which is
 folded and takes the form of the initial open string.

Formally, the worldsheet orientifold represents a doubling of the
worldsheet with the orientation reversal on the second sheet. The
fundamental domain $[0,\pi]$ is extended to $\Sigma=[0,2\pi]$ with
formation of folds at the ends of the interval $[0,\pi]$.
\section{Solutions of the e.m. field equations}
Field equations for Einstein-Maxwell system in the Kerr-Schild class
were obtained in \cite{DKS}.
We will concentrate here on the electromagnetic excitations of the
Kerr ring. These e.m. fields are aligned to the Kerr PNC and
described by the tetrad components of the self-dual tensor
$ \cF _{12} =AZ^2, \qquad\cF _{31} =\gamma Z - (AZ),_1  $.
The field equations are
\be A,_2 - 2 Z^{-1} \cZ Y,_3 A  = 0 , \label{3}\ee
\be \cD A+  \cZ ^{-1} \gamma ,_2 - Z^{-1} Y,_3 \gamma =0 .
\label{4}\ee
For the sake of simplicity we consider the gravitational
Kerr-Schild field as stationary.
The corresponding oscillating solutions can be obtained by
introduction of a complex retarded time
parameter $\t = t_0 +i\sigma = \t |_L$ which is determined as a result of the
intersection of the left (L) null plane
and the complex world line (for details see \cite{Bur-nst}).
The left null planes are the left generators of the complex null cones
and play the role of null cones in the complex retarded-time construction.
The $\t$ parameter satisfies to relations
$(\t),_2 =(\t),_4 = 0 $.
It allows one to represent the equation (\ref{3}) in the form
$(AP^2),_2=0 $, and to get the following general solution
\be A= \psi(Y,\t)/P^2.
\label{A}\ee
Action of the operator $\cD$ on the variables
$Y, \bar Y $ and $ t_0$ is following
$ \cD Y = \cD \bar Y = 0,\qquad
\cD t_0 = P^{-1} $.
The equation (\ref{4}) takes the form
$\d_{t_0} A = -(\gamma P),_{\bar Y} $.
For stationary background $P=2^{-1/2}(1+Y\bar Y)$, and
$\dot P = 0$.  The coordinates $Y$,  and $\t$ are independent from
$\bar Y$, which allows us to integrate this equation and
we obtain the following general solution

$\gamma  = - P^{-1}\int \dot A d\bar Y =
 - P^{-1}\dot \psi (Y,\t) \int  P^{-2}d\bar Y  =
 \frac{2^{1/2}\dot \psi} {P^2 Y} +\phi (Y,\t)/P $,
where $\phi$ is an arbitrary analytic function of $Y$ and $\t$.

The term $\gamma$  in
$ \cF _{31} =\gamma Z - (AZ),_1  \ $
describes a part of the null
electromagnetic radiation   which
falls of asymptotically as $1/r$ and
propagates along the Kerr principal null congruence $e^3$.
As it was discussed in \cite{Bur-nst,Ori}, this field acquires
interpretation of the vacuum zero point field
 which has to be regularized similar to the zero point field in the
Casimir effect.

\section{Axial strings} 
Considering the second term in $ \cF _{31}$,
we obtain the terms containing the factors which depend on coordinate
$Y = e^{i\phi} \tan \frac {\theta} 2 $ and turn out to be
singular at the axis of symmetry (z-axis).

These factors result the appearance of
two  half-infinite lines of singularity,  $z^+$ and $z^-$,
which correspond to $\theta =0, \quad Y\to 0$ and
$\theta=\pi, \quad Y\to \infty$ and coincide
with corresponding axial lightlike rays of the Kerr PNC.
On the ``positive'' sheet of the Kerr background these two
half-rays are directed outward. They are going
from the ``negative" sheet and  appear on the ``positive'' sheet
passing through the Kerr ring (see Fig. 2).

\begin{figure}[ht]
\centerline{\epsfig{figure=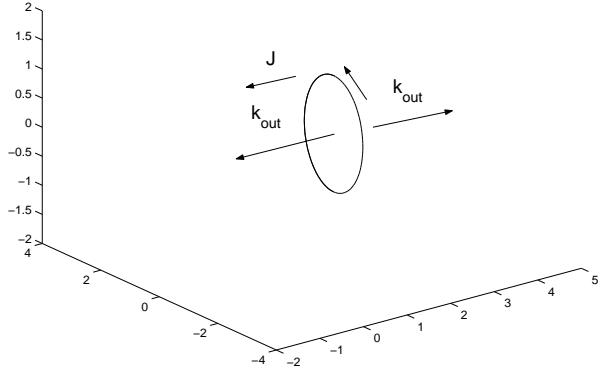,height=6cm,width=8cm}}
\caption{Schematic description of the Kerr spinning particle.
Circular string and two axial half-infinite strings directed outwards.}
\end{figure}
We will be interested in the wave terms and omit the terms describing the
longitudinal components and the field $\gamma$.
By using the expressions for the Kerr tetrad
we obtain that the wave terms
 of the e.m. field take the form\fn{Here the null coordinates are used
$2^{1\over2}\z = x+iy ,\quad 2^{1\over2} \Z = x-iy , \quad
2^{1\over2}u = z - t ,\quad 2^{1\over2}v = z + t .$}
\be \cF |_{wave} =f_R \ d \z \wedge d u  +
f_L \ d \Z \wedge d v ,
\label{cFLR}
\ee

where the factor $f_R = (AZ),_1$
describes the ``right"  waves propagating along the $z^+$ half-axis,
and the factor
$f_L =2Y \psi (Z/P)^2 + Y^2 (AZ),_1$
describes the ``left"  waves propagating along the  $z^-$ half-axis.

Since $Z/P=(r+ia \cos \theta)^{-1}$, all
the terms are also singular at the Kerr ring $r=\cos \theta =0$.
Therefore, the singular excitations of the Kerr ring turn out to be
connected with the axial singular waves.

Traveling waves along the Kerr ring are generated by the function
$\psi _n (Y,\t) = q Y^n \exp {i\omega _n \t}
\equiv q (\tan \frac \theta 2)^n \exp {i(n\phi + \omega _n \t)}$
which has near the Kerr ring the form $\psi =\exp {i(n\phi + \omega t)}$, and
index $|n|$ corresponds to the number of the wave lengths along the Kerr ring.
The factor $Y^n$ leads to the appearance of the $z^\pm$ axial singularities.
When considering asymptotical properties
of these singularities by $r \to \infty $, we have
 $z=r\cos \theta$, and for the distance $\rho$ from the $z^+$ axis we have
the expression $\rho = z \tan \theta \simeq 2 r |Y| $ by $Y\to 0$.
Therefore, for the asymptotical region near the $z^+$ axis we have to put
$Y = e^{i\phi} \tan {\frac \theta 2} \simeq  e^{i\phi} \frac \rho {2r}$, and
$|Y|\to 0$,
while for the asymptotical region near the $z^-$ axis
$Y = e^{i\phi} \tan {\frac \theta 2 } \simeq  e^{i\phi} \frac {2r} \rho  $,
and $|Y|\to \infty$.  The parameter $\t=t -r -ia \cos \theta$ takes near the
z-axis the values $\t _+ = \t |_{z^+}= t-z-ia$, and $ \t _- = \t |_{z^-}
=t+z +ia$.

For $|n|>1$ the solutions contain the axial
singularities which do not fall of  asymptotically and are increasing.
Therefore, the treatment has to be restricted by the cases $|n|\le 1$.
The leading singular wave for $n=1$, $\cF^-_1$,
propagates to $z=-\infty$ along the $z^-$ half-axis, and
the leading singular wave for $n=-1$, $\cF^+_-1$
propagates to $z=+\infty$ along the $z^+$ half-axis.
They have the form
\be
\cF^-_1=\frac {4q e^{i2\phi +i\omega _1 \t _-}}
{\rho ^2} \ d \Z \wedge dv, \qquad
\cF^+_{-1}= -
\frac {4q e^{-i2\phi+i\omega _{-1} \t _+}} {\rho ^2} \ d \z \wedge du ,
\ee
The waves with $n=0$, $\cF^\pm_0$, are regular.

One sees that the partial solutions turns out to be asymmetric with respect
to the $z^{\pm}$ half-axis,
which can lead to a nonstationarity via a recoil.
This e.m. field can also be obtained from the potential
$\cA = - AZ e^3 - \chi d \Y $,
where $A=\psi /P^2$ and $\chi = \int P^{-2} \psi d Y $, and
$\Y$ being kept constant in this integration.

Each of the partial solutions
represents the singular plane-fronted e.m. wave propagating along
$z^+$ or $z^-$ half-axis without damping.
It is easy to point out the corresponding
self-consistent solution of the Einstein-Maxwell field equations which
belongs to the well known class of pp-waves \cite{KraSte}.
These singular pp-waves propagate outward
along the $z^+$ and/or  $z^-$ half-axis and can be regularized by
a Higgs field leading to the axial stringy currents.

By the analysis of the corresponding gravitational field equations, one can
see
that the resulting metric acquires an imaginary contribution to the mass term,
which is the evidence that the NUT parameter has to be involved, and also that
the metric is going out of the Kerr-Schild class.
Therefore, the obtained singular strings are related to the Dirac monopole
string.

To exclude the monopole charge and to get a symmetric stringy solution,
one has to
use a combination of the $n=\pm 1$ excitations. There is one solution
containing combination of three terms with $n=-1, 0, 1$ with
$\omega _1 = -\omega _{-1}=\omega$ and $\omega _0=0$. It represents
especial interest since it has an electric charge and a smooth
 e.m. field having one half of the wavelength packed along the Kerr string.

Note, that orientifold structure of the Kerr circular string
admits apparently the excitations with $n= \pm 1/2$ too, so far as the
negative half-wave can be packed on the covering space turning into
positive one on the second sheet of the orientifold. However,
this question demands an additional consideration.
The above axial half-infinite strings are carriers of pp-waves
and for the moving particles these pp-waves are modulated by de Broglie
periodicity. It allows one to conjecture that this string could be a
carrier of the wave function.

At first sight these half-infinite strings looks very strange.
However, they are half-infinite (like the Dirac monopole string)
only for the free particles. By interaction they can form
the closed quantized loops. Such a sort of the linked strings was
considered by Garriga and Vachaspati \cite{GarVac}.
For the field description of coupled strings
the Witten field model of superconducting strings \cite{Wit} has
to be used. The stringy Kerr source model opens an interesting way for
topological realizations of the standard model of particle physics and
gives a new view of some quantum phenomena.
Application of the supersymmetric version of the Witten field model
to the Kerr baglike source was discussed in \cite{Bag}.

\section*{ Acknowledgments}
We are thankful to organizers of the Workshop
SQS'03 for financial support and also to many participants for very
useful discussions, in particular to A. Zheltukhin, I. Bandos, A.
Tseytlin and G. Alekseev.

\end{document}